\documentclass[12pt]{article}
\usepackage{graphicx}
\usepackage{lscape}
\usepackage{citesort}

\begin{document}

\def\lra{\stackrel{\leftrightarrow}}
\def\ds{\displaystyle}
\def\beq{\begin{equation}}
\def\eeq{\end{equation}}
\def\bea{\begin{eqnarray}}
\def\eea{\end{eqnarray}}
\def\beeq{\begin{eqnarray}}
\def\eeeq{\end{eqnarray}}
\def\ve{\vert}
\def\vel{\left|}
\def\ver{\right|}
\def\nnb{\nonumber}
\def\ga{\left(}
\def\dr{\right)}
\def\aga{\left\{}
\def\adr{\right\}}
\def\lla{\left<}
\def\rra{\right>}
\def\rar{\rightarrow}
\def\nnb{\nonumber}
\def\la{\langle}
\def\ra{\rangle}
\def\ba{\begin{array}}
\def\ea{\end{array}}
\def\tr{\mbox{Tr}}
\def\ssp{{\Sigma^{*+}}}
\def\sso{{\Sigma^{*0}}}
\def\ssm{{\Sigma^{*-}}}
\def\xis0{{\Xi^{*0}}}
\def\xism{{\Xi^{*-}}}
\def\qs{\la \bar s s \ra}
\def\qu{\la \bar u u \ra}
\def\qd{\la \bar d d \ra}
\def\qq{\la \bar q q \ra}
\def\gGgG{\la g^2 G^2 \ra}
\def\q{\gamma_5 \not\!q}
\def\x{\gamma_5 \not\!x}
\def\g5{\gamma_5}
\def\sb{S_Q^{cf}}
\def\sd{S_d^{be}}
\def\su{S_u^{ad}}
\def\ss{S_s^{??}}
\def\sbp{{S}_Q^{'cf}}
\def\sdp{{S}_d^{'be}}
\def\sup{{S}_u^{'ad}}
\def\ssp{{S}_s^{'??}}
\def\sig{\sigma_{\mu \nu} \gamma_5 p^\mu q^\nu}
\def\fo{f_0(\frac{s_0}{M^2})}
\def\ffi{f_1(\frac{s_0}{M^2})}
\def\fii{f_2(\frac{s_0}{M^2})}
\def\O{{\cal O}}
\def\sl{{\Sigma^0 \Lambda}}
\def\es{\!\!\! &=& \!\!\!}
\def\ap{\!\!\! &\approx& \!\!\!}
\def\ar{&+& \!\!\!}
\def\ek{&-& \!\!\!}
\def\kek{\!\!\!&-& \!\!\!}
\def\cp{&\times& \!\!\!}
\def\se{\!\!\! &\simeq& \!\!\!}
\def\eqv{&\equiv& \!\!\!}
\def\kpm{&\pm& \!\!\!}
\def\kmp{&\mp& \!\!\!}
\renewcommand{\thefootnote}{\fnsymbol{footnote}}


\def\simlt{\stackrel{<}{{}_\sim}}
\def\simgt{\stackrel{>}{{}_\sim}}


\title{
         {\Large
                 {\bf
Unparticle effects in rare $t \rar c g g$ decay 
                 }
         }
      }

\author{\vspace{1cm}\\
{\small T. M. Aliev$^a$ \thanks
{e-mail: taliev@metu.edu.tr}~\footnote{Permanent Address: 
Institute of Physics, Baku, Azerbaijan}\,\,,
A. Bekmezci$^b$ \thanks
{e-mail: bekmezci@mu.edu.tr}\,\,,
M. Savc{\i}$^a$ \thanks
{e-mail: savci@metu.edu.tr}} \\
{\small (a) Physics Department, Middle East Technical University,
06531 Ankara, Turkey}\\
{\small (b)  Physics Department, Mu\u{g}la University, Mu\u{g}la, Turkey} }

\date{}

\begin{titlepage}
\maketitle
\thispagestyle{empty}

\begin{abstract}
Rare $t \rar c g g$ decay can only appear at
loop level in the Standard Model (SM), and naturally they are strongly suppressed.
These flavor changing decays induced by the mediation of spin--0 and spin--2 
unparticles, can appear at tree level in unparticle physics. In this work
the virtual effects of unparticle physics in the flavor--changing 
$t \rar c g g$ decay is studied. Using the
SM result for the branching ratio of the $t \rar c g g$ decay, 
the parameter space of $d_{\cal U}$ and $\Lambda_{\cal U}$,
where the branching ratio of this decay exceeds the one predicted by the
SM, is obtained. Measurement of the branching ratio larger 
than $10^{-9}$ can give valuable information for establishing unparticle 
physics.
\end{abstract}

\end{titlepage}

\section{Introduction}

Rare decays, induced by the flavor--changing neutral current (FCNC)
transitions, is quite a promising research area and attracts theoretical
interest as a potential testing ground for checking predictions
of the Standard model (SM) at loop level, as well as search for new physics
effects (NP) beyond the SM. The effects of NP in rare decays can appear in
two different ways: i) via new contributions to the Wilson coefficients 
existing in the SM,  ii) via appearance of new operators with new Wilson
coefficients which are absent in the present SM.

The impressive and exciting results on the FCNC decays in B--meson sector
were observed at two B--meson factories KEK and BELLE
\cite{R9101,R9102,R9103}, and at CLEO \cite{R9104}, which are in good 
agreement with the SM prediction. 

The interest to the study of FCNC decays in t--quark sector can be
explained on the following reasons: i) In many models beyond the SM the 
new physics scale is closer to the t--quark mass, and ii) many two--body
t--quark FCNC decays, like $t \rar c V ~(V = g,\gamma,Z)$ and $t \rar c
H$ are highly suppressed in the SM due to the GIM mechanism and their
branching ratios are of the order $10^{-11}\div 10^{-14}$ \cite{R9105,R9106}.
These branching ratios are practically impossible to measure at LHC
\cite{R9107} or at International Linear Collider (ILC) \cite{R9108}. But
many models of NP predict that the branching ratios of the above--mentioned
FCNC decays are much larger compared to that obtained in the SM (\cite{R9109}
and references therein).

The t--quark three--body FCNC decays like $t \rar cWW,cZZ,bWZ$ are also
discussed in the framework of SM \cite{R9110,R9111,R9112} and its beyond
\cite{R9113}. It is shown in \cite{R9110,R9111,R9112} that the rate of higher order
three--body FCNC decay $t \rar cgg$ exceeds the rate of lower order $t \rar
cgg$  decay. 

As has already been noted, FCNC processes are very sensitive to the new
physics effects. One such model is the so--called unparticles is recently
proposed by H. Georgi \cite{R9114}. The main idea in this model is that 
at very high energies the SM fields and the Bank--Zaks (BZ) fields with a
nontrivial infrared fixed point interact. The interaction between these two
sectors is due to the exchange of particles with a large mass scale 
$M_{\cal U}$. The interaction below this scale is nonrenormalizable and is
suppressed by a power of $M_{\cal U}$. The renormalizable couplings of BZ
fields then produce the dimensional transmutation and the scale invariant
unparticle emerges below the scale $\Lambda_{\cal U}$, and the unparticle
stuff with scale dimension $d_{\cal U}$, looks like massless invisible
particles with noninteger number $d_{\cal U}$. For this reason, production
of unparticles might be detectable in missing energy processes.
Phenomenology of unparticle physics is studied extensively in the 
literature
\cite{R9115,R9116,R9117,R9118,R9119,R9120,R9121,R9122,R9123,R9124,R9125,R9126,R9127}. 
In the present work we study $t \rar cgg$ 
($t \rar c\gamma \gamma$) decay in the framework of an unparticle physics.
Organization of this paper is as follows: In section--2 calculation of the
$t \rar cgg$ and $t \rar c\gamma \gamma$ decays are presented. In section--3
numerical results and discussion are given.            

\section{Formalism}

In this section we calculate the branching ratio of $t \rar cgg$
($t \rar c\gamma \gamma$) decay in unparticle physics. As we have noted
already, below $\Lambda_{\cal U}=1~TeV$ the interaction between SM fields
and BZ fields become an effective operator, i.e., it has the following form:
\bea
\label{e9101}
{\cal L}_{int} = {1\over \Lambda_{\cal U}} O_{SM} O_{\cal U}~.
\eea
Obviously, High--dimension operators should be suppressed by inverse power
of $\Lambda_{\cal U}$. Therefore, we should choose the appropriate operators
with the lowest dimension. Also, the effective interaction should satisfy
the SM gauge symmetry. The effective Lagrangian of scalar and tensor
unparticle operators with SM fields are given in \cite{R9128}:

a) scalar unparticle

\bea
\label{e9102}
&&\lambda_0 {1 \over \Lambda_{\cal U}^{d_{\cal U}-1}} \bar{f}f O_{\cal U}~,\nnb \\
&&\lambda_0 {1 \over \Lambda_{\cal U}^{d_{\cal U}-1}} \bar{f}i\gamma_5 f 
O_{\cal U}~,\nnb \\
&&\lambda_0 {1 \over \Lambda_{\cal U}^{d_{\cal U}}} \bar{f}\gamma_\mu(\gamma_5) f 
\partial_\mu O_{\cal U}~,\nnb \\
&&{1 \over \Lambda_{\cal U}^{d_{\cal U}}} \Big[
\lambda_0 G_{\alpha\beta} G^{\alpha\beta} + \lambda_0^\prime G_{\alpha\beta}
\widetilde{G}^{\alpha\beta}\Big] O_{\cal U}~,
\eea

b) tensor unparticle

\bea
\label{e9103}
&& -{1\over 4} \lambda_2 {1 \over \Lambda_{\cal U}^{d_{\cal U}}} \bar{\psi} i 
\Big[\gamma_\mu (\gamma_5) \lra{\cal D}_\nu + \gamma_\nu \gamma_5
\lra{\cal D}_\mu \Big] \psi O_{\cal U}^{\mu\nu}~,\nnb \\
&& {1 \over \Lambda_{\cal U}^{d_{\cal U}}} \Big[ \lambda_2 G_{\mu\nu}
G_\nu^\alpha + \lambda_2^\prime G_{\mu\alpha} \tilde{G}_\nu^\alpha \Big]
O_{\cal U}^{\mu\nu}~,
\eea
where 
\bea
{\cal D}_\mu =\partial_\mu + i g {\tau^a \over 2} W_\mu^a + i
g^\prime {V\over 2} B_\mu~,\nnb
\eea 
is the covariant derivative in the SM, $G^{\alpha\beta}$ is the gauge field 
strength tensor (gluon, photon, as well as weak gauge bosons), 
\bea
\widetilde{G}_{\alpha\beta} = {1\over 2} \varepsilon_{\mu\nu\alpha\beta}
G^{\alpha\beta}~,\nnb
\eea
$f$ is a standard model fermion, and $\psi$ stands for a SM doublet or
singlet fermion, and $\lambda_i$ are the dimensionless effective couplings.
Note that we will neglect the third term in scalar unparticle case because
this term contain an extra $1/\Lambda_{\cal U}$ factor.

It follows from Eqs. (\ref{e9102}) and (\ref{e9103}) that the flavor
violating $t \rar c g g$, $t \rar c \gamma \gamma$ decays can take place at
tree level in unparticle physics, while they exist at loop level in the SM, 
and this is the main reason why we consider them in unparticle physics.    

Scale invariance determines the form of the propagators within normalization
factor. The propagators corresponding to scalar and tensor unparticles are
\bea
\label{e9104}
{\cal D}(q^2) \es {A_{d_{\cal U}} \over 2 \sin (d_{\cal U}\pi)}
(-q^2)^{d_{\cal U}-2}~,~~~\mbox{\rm and},\\
\label{e9105}
\Delta_{\mu\nu\rho\sigma} \es {\cal D} (q^2) {\cal P}_{\mu\nu\rho\sigma}~.
\eea
For the transverse and traceless tensor operators $O_{\mu\nu}$ \cite{e9128}
we have
\bea
\label{e9106}
{\cal P}_{\mu\nu\rho\sigma} \es {1\over 2} \Bigg\{ \Pi_{\mu\nu}
\Pi_{\rho\sigma}  + \Pi_{\mu\sigma} \Pi_{\nu\rho} - {2\over 3}
\Pi{\mu\rho}  + \Pi_{\nu\sigma} \Bigg\}~,
\eea 
while in conformal filed theories \cite{R9129}
\bea
\label{e9107}
T_{\mu\nu\rho\sigma} \es {1\over 2} \Bigg[(g_{\mu\rho} g_{\nu\sigma} +
g_{\mu\sigma} g_{\nu\rho}) + {4 - d_{\cal U} (d_{\cal U}+1) \over 
2 d_{\cal U} (d_{\cal U}-1)} g_{\mu\nu} g_{\rho\sigma} \nnb \\ 
\ek
2 \Bigg( {d_{\cal U} -2 \over d_{\cal U}} \Bigg)
\Bigg(g_{\mu\rho} {k_\nu k_\sigma\over k^2} + g_{\mu\sigma} 
{k_\nu k_\rho\over k^2} +
g_{\nu\rho} {k_\mu k_\sigma\over k^2} + g_{\nu\sigma} 
{k_\mu k_\rho\over k^2} \Bigg) \nnb \\
\ar 4 {d_{\cal U} -2 \over d_{\cal U} (d_{\cal U} -1)}  
\Bigg(g_{\mu\nu} {k_\rho k_\sigma\over k^2} + g_{\rho\sigma} 
{k_\mu k_\nu\over k^2}\Bigg) +
8{(d_{\cal U} - 2) (d_{\cal U} - 3) k_\mu k_\nu k_\rho k_\sigma
\over d_{\cal U} (d_{\cal U} - 1) (k^2)^2 } \Bigg]~,
\eea
where
\bea
\label{e9108}
\Pi_{\mu\nu} = - g_{\mu\nu} + a {\ds q_\mu q_\nu \over \ds q^2}~,
\eea
where
\bea
a = \left\{ \begin{array}{l}
1~,~~\mbox{\rm for transverse vector operator, and}~,\\ \\
{\ds 2 (d_{\cal U} - 2) \over \ds (d_{\cal U} - 1)} ~,~~\mbox{\rm in
conformal field theory}~.
\end{array} \right. \nnb
\eea
In the present work we follow the Georgi's approach \cite{R9114}, namely,
Feynman propagators of the unparticle operator $O_{\cal U}$ is determined by
the scalar invariance.

The factor $A_{\cal U}$ in Eqs. (\ref{e9104}) and (\ref{e9105}) is
\bea
A_{\cal U} = {16 \pi^{5/2} \over (2\pi)^{2 d_{\cal U}}}
{\Gamma(d_{\cal U} + 1/2) \over \Gamma(d_{\cal U} - 1) 
\Gamma(2d_{\cal U})}~. \nnb
\eea

In order to calculate the decay rate $t \rar c g g$ and $t \rar c \gamma
\gamma$ decays we also need $g g {\cal U}$, $\gamma \gamma {\cal U}$ and
fermion--fermion unparticle interaction vertices. From Eqs. (\ref{e9102}) 
and (\ref{e9103}) we get the following expressions for the above--mentioned
vertices:

a) fermion--fermion scalar unparticle 
\bea
{\lambda_0 \over \Lambda_{\cal U}^{d_{\cal U} - 1} } \bar{c} \Big[ C_S +
i\gamma_5 C_P \Big] t~, \nnb
\eea

b) gluon--gluon scalar unparticle 
\bea
{\lambda_0 \over \Lambda_{\cal U}^{d_{\cal U}}} 2 \Bigg\{ {\lambda^a \over
2}{\lambda^b \over 2} \Bigg\} \Big[ (k_1 \cdot k_2) g_{\mu\nu} - k_{1\mu} \cdot
k_{2\nu} - 2 \epsilon_{\mu\nu\alpha\beta} k_{1\alpha} k_{2\beta} \Big]
\varepsilon_\mu^a(k_1) \varepsilon_\nu^b(k_2)~,\nnb
\eea

Photon--photon scalar unparticle vertex can be obtained from gluon--gluon scalar
unparticle by making the replacement
\bea
\Bigg\{ {\lambda^a \over 2}{\lambda^b \over 2} \Bigg\} \rar 1~,\nnb
\eea  
and omitting color indices in $\varepsilon_\mu^a$, and hence we get:

c) fermion--fermion tensor unparticle
\bea
{1 \over 4 \Lambda_{\cal U}^{d_{\cal U}}} \Big\{ \lambda_2 \Big[ \gamma_\mu
(p_c + p_t)_\nu + \gamma_\nu (p_c + p_t)_\mu \Big] + 
\lambda_2^\prime \Big[ \gamma_\mu \gamma_5 (p_c + p_t)_\nu + \gamma_\nu
\gamma_5(p_c + p_t)_\mu \Big] \Big\}~, \nnb
\eea
 
d) gluon--gluon tensor unparticle 
\bea
{1 \over \Lambda_{\cal U}^{d_{\cal U}}}  \Bigg[\lambda_2 \Bigg( 
\Bigg\{ {\lambda^a \over 2} {\lambda^b \over 2}\Bigg\} 
K_{\mu\nu\rho\sigma}^S \ar \Bigg[ {\lambda^a \over 2} 
{\lambda^b \over 2}\Bigg] K_{\mu\nu\rho\sigma}^A \Bigg) \nnb \\
\ar \lambda_2^\prime \Bigg( 
\Bigg\{ {\lambda^a \over 2} {\lambda^b \over 2}\Bigg\} 
F_{\mu\nu\rho\sigma}^S + \Bigg[ {\lambda^a \over 2} 
{\lambda^b \over 2}\Bigg] F_{\mu\nu\rho\sigma}^A \Bigg)
\Bigg] \varepsilon_\mu^a(k_1) \varepsilon_\nu^b(k_2)~,\nnb 
\eea
where
\bea
K_{\mu\nu\rho\sigma}^{S(A)} \es {1 \over 2} \Big\{(k_1\cdot k_2) g_{\mu\rho}
g_{\nu\sigma} + g_{\mu\nu} k_{1\rho} k_{2\sigma} -
g_{\nu\sigma} + g_{\mu\nu} k_{1\rho} k_{2\mu} - 
g_{\mu\rho} + g_{\mu\nu} k_{1\nu} k_{2\sigma} \nnb \\
\kpm \Big[
(k_1\cdot k_2) g_{\mu\nu}
g_{\rho\sigma} + g_{\mu\rho} k_{1\nu} k_{2\sigma} -
g_{\rho\sigma} + g_{\mu\rho} k_{1\nu} k_{2\mu} -
g_{\mu\nu} + g_{\mu\rho} k_{1\rho} k_{2\sigma} \Big] \Big\}~,\nnb \\ \nnb \\
K_{\mu\nu\rho\sigma}^{S(A)} \es {1 \over 2} \Big(k_{1\rho} k_{2\beta}
\epsilon_{\mu\nu\beta\sigma} - k_{1\alpha} k_{2\beta} g_{\mu\rho}
\epsilon_{\sigma\alpha\beta\nu} \mp k_{1\beta} k_{2\rho}
\epsilon_{\mu\nu\beta\sigma} \mp k_{1\beta} k_{2\alpha} g_{\rho\nu}
\epsilon_{\sigma\alpha\beta\mu}\Big)~.\nnb
\eea

Photon--photon tensor unparticle vertex can easily be obtained from
gluon--gluon tensor unparticle vertex by making the following replacements:
\bea
\Bigg\{ {\lambda^a \over 2} {\lambda^b \over 2} \Bigg\} \rar 1~,~~~
\Bigg[ {\lambda^a \over 2} {\lambda^b \over 2} \Bigg] \rar 0~,\nnb
\eea
and omitting color indices in $\varepsilon_\mu^a$.

Now we are ready to calculate the branching ratio of the $t\rar c g g$ and 
$t \rar c \gamma \gamma$ decays. In calculation of the branching ratios of
these decays there appear infrared and collinear divergences. There are
three possible sources of these singularities:

1) One gluon (photon) flying parallel to the c--quark,

2) two gluons (photons) flying parallel to each other, and,

3) one of the gluons (photons) is soft.

First and second cases are related with the collinear singularity, while the
last case is related to the infrared singularity.

In order to avoid the singularity in case--1, it is enough to take into
account mass of the c--quark in calculations. There are two different ways
to prevent the singularities in cases--2 and --3, one of them is to put
cut--off factor in ``dangerous'' integration limit where
singularities are present (see \cite{R9130}).

Using the Feynman rules for the matrix element of the $t \rar c g g $ decay
exchanging the scalar and tensor unparticles, we get respectively,
\bea
M_S \es T_{\mu\nu}^+ \Bigg\{ {\lambda^a \over 2} {\lambda^b \over 2} \Bigg\}
\bar{c} \Big[ C_S + C_P \gamma_5 \Big] t \varepsilon_\mu^a(k_1)
\varepsilon_\nu^b(k_2)~, \nnb \\ \nnb \\
M_T \es \Bigg( T_{\mu\nu\rho\sigma}^+ \Bigg\{ {\lambda^a \over 2} {\lambda^b \over
2} \Bigg\} + T_{\mu\nu\rho\sigma}^- \Bigg[ {\lambda^a \over 2} {\lambda^b
\over 2} \Bigg] \Bigg)
{\cal P}_{\rho_1,\sigma_1,\rho\sigma}  \bar{c} \Big\{ \lambda_2 \Big[
\gamma_{\rho_1} (p_c+p_t)_{\sigma_1} + 
\gamma_{\sigma_1} (p_c+p_t)_{\rho_1} \nnb \\
\ar \lambda_2^\prime \Big[ \gamma_{\rho_1} \gamma_5 (p_c+p_t)_{\sigma_1} + 
\gamma_{\sigma_1} \gamma_5 (p_c+p_t)_{\rho_1} \Big] \Big\} t
\varepsilon_\mu^a \varepsilon_\nu^b~, \nnb
\eea
where
\bea
T_{\mu\nu}^+ \es {\lambda_0 \over \Lambda_{\cal U}^{2 d_{\cal U}-1}} {A_{\cal U}
\over \sin (d_{\cal U}\pi) } {1 \over (q^2)^{2-d_{\cal U}}}
\Big\{ \lambda_0 [k_{1\nu} k_{2\mu} - g_{\mu\nu} (k_1 \cdot k_2)] +
\lambda_0^\prime \epsilon_{\mu\nu\alpha\beta} k_{1\alpha} k_{2\beta} \Big\}~,
\nnb \\ \nnb \\
T_{\mu\nu\rho\sigma}^{\pm} \es 
{1 \over 4 \Lambda_{\cal U}^{2 d_{\cal U}}} {A_{\cal U} 
\over 2 \sin (d_{\cal U}\pi) } {1 \over (q^2)^{2-d_{\cal U}}}
\Big( \lambda_2 K_{\mu\nu\rho\sigma}^{S(A)} + 
\lambda_2^\prime F_{\mu\nu\rho\sigma}^{S(A)} \Big)~.
\eea

In further analysis we take into account the following fact. It is well
known that \cite{R9131,R9132} if in the problem under consideration there
appear two or more external gluons whose polarization sum is $\sum_\lambda
\varepsilon_\mu^\ast (k,\lambda) \varepsilon_\nu (k,\lambda) = -
g_{\mu\nu}$, gauge invariance is violated. In our calculation we
choose the following expression for the polarization sum of the gluons,
simultaneously which are transverse to massless vector boson momenta $k_1$
and $k_2$,
\bea
P_{\mu\nu} \es \sum_{\lambda=1,2} \varepsilon_\mu^\ast (k,\lambda)
\varepsilon_\nu (k,\lambda)~, \nnb \\
\es - g_{\mu\nu} + { k_{1\mu} k_{2\nu} +  k_{1\nu} k_{2\mu} \over 
k_1\cdot k_2}~, \nnb
\eea
which leads to the gauge invariant result for on--shell massless vector
mesons.

Using the matrix element for the $t \rar c x x$ $(x=g,\gamma)$ decay, in the
rest frame system of the decaying t--quark, we get for the differential
decay width
\bea
d\Gamma = {1 \over 256 m_t \pi^3} C_X \vel M_X^\prime \ver^2 dE_C \, dE_1~, \nnb
\eea
where prime means summation over gluon (photon) is performed, and $C_X$ is
the color factor whose values are presented in the table.

\newcommand{\rb}[1]{\raisebox{3.0ex}[0pt]{#1}}
\begin{table}[h]
\renewcommand{\arraystretch}{2.0}
\addtolength{\arraycolsep}{2.5pt}
$$
\begin{array}{|c|c|c|c|}
\hline 
                                            &
\multicolumn{2}{|c|}{t \rar c g g}         &                             \\ \cline{2-3}
               & \mbox{\rm Antisymmetric}    & \mbox{\rm Symmetric}   & 
\raisebox{3.5ex}[0pt]{$t \rar c \gamma \gamma$} \\ \hline
               & \frac{N^2-1}{2}  & \frac{N^2-1}{2 N^2}~~\mbox{\rm (Singlet)} &               \\
\rb{$C_X$}  &          &  \frac{(N^2-1)(N^2-2)}{2 N^2}~~\mbox{\rm (Adjoint)}  &                       
\rb{1}  \\ \hline
\end{array}
$$
\caption{}
\renewcommand{\arraystretch}{1}
\addtolength{\arraycolsep}{-5pt}
\end{table} 

In order to calculate the branching ratio, we take into account that the 
$t \rar b W$ decay is the dominant channel of the t--quark, and use
$\Gamma(t \rar b W) = 1.55~GeV$.

\section{Numerical analysis}

In this section we study the sensitivity of the branching ratio on the
scaling dimension parameter $d_{\cal U}$, energy scale $\Lambda_{\cal U}$ 
and the coupling constants. In numerical analysis we choose the scaling
dimension $d_{\cal U}$ in the range $1 < d_{\cal U} < 2$. The main reason
for choosing $d_{\cal U} > 1$ is that in this region the decay rate is free
from the nonintegrable singularity \cite{R9114}. As has already been mentioned, 
there appear singularities for $d_{\cal U} > 2$. Therefore we will consider
the above--mentioned restricted domain of $d_{\cal U}$. The values of the 
off--diagonal t--c unparticle coupling constants $C_S$ and $C_P$ are chosen
in the range $10^{-1} \div 10^{-3}$. For the parameter alpha we choose three
different values $\alpha = 0.1;~0.5;~1.0$. Note that the branching ratio of
the $t \rar c g g$ decay in the SM is calculated in \cite{R9109} which
predicts ${\cal B}(t \rar c g g)\simeq 1.02\times 10^{-9}$, when the
cut--off parameter $C$ is taken $C=10^{-3}$. Our numerical calculations
shows that when the cut--off parameter $C$ varies in the range $c=0.001 \div
0.1$ for a given set of the fixed values of $C_P$ and $C_S$, no substantial
change in the value of the branching ratio is observed, the variation being
about three times. The above--mentioned value of the branching ratio of the
$t \rar c g g$ decay in the SM is too small to be observable in the
forthcoming LHC experiments. For this reason any experimental observation of
the $t \rar c g g$ decay will definitely indicate the appearance of the new
physics beyond the SM. Therefore, the observability limit of the 
$t \rar c g g$ decay can be assumed to be ${\cal B}(t \rar c g g) = 10^{-9}$.
In this connection there follows the question about the range of values of 
$d_{\cal U}$ for which the branching ratio is larger than $10^{-9}$, at the
value $\Lambda_{\cal U}=1~TeV$ of the cut--off parameter and at fixed values
of the effective couplings $C_P$ and $C_S$ (in the presence of the scalar
unparticle operator). 

Our numerical analysis predicts the following results:

\begin{itemize}
\item at $C_S=C_P=10^{-1}$, $d_{\cal U} < 1.5~(<1.53,~<1.55)$, and when  
$C=0.1(10^{-2},~10^{-3})$;

\item at $C_S=C_P=10^{-2}$, $d_{\cal U} < 1.2~(<1.24,~<1.25)$, and when  
$C=0.1(10^{-2},~10^{-3})$;

\item at $C_S=C_P=10^{-3}$,                                     

\end{itemize}
the corresponding branching ratios are larger compared to the the SM 
result.

It follows from the above--presented results that the restrictions to the
values of $d_{\cal U}$ in both decays, for which the branching ratio exceeds 
$10^{-9}$, are practically the same.

As an illustration of our analysis, we present in Fig. (1) the dependence of
the branching ratio of the $t \rar c g g$ decay on $d_{\cal U}$, at
$C_S=C_P=10^{-2}$, $C=10^{-2}$, when scalar unparticle operator is the
mediator. Here the parameter $\alpha$ is defined as $\alpha =
\lambda_0^\prime/\lambda_0$, and we set $\lambda_0=1$. From this figure we
see that up to $d_{\cal U}=1.1$ the perpendicular spin polarization exceeds
the parallel spin polarization for two--gluon system at $\alpha=1$.

These results are quite interesting since they give valuable information
about the scaling parameter $d_{\cal U}$, as well as information about
gluon--gluon unparticle coupling constants.   

For the tensor operator case we obtain the restrictions 
$d_{\cal U} < 1.4~(<1.55,~<1.58)$ at $C=0.1(10^{-2},~10^{-3})$ for the 
$t \rar c g g$ decay, for which the branching ratio exceeds $10^{-9}$, at
$\Lambda_{\cal U} = 1~TeV$.

Depicted in Fig. (2) is the dependence of the branching ratio for the $t
\rar c g g$ decay on $d_{\cal U}$ at $\Lambda_{\cal U} = 1~TeV$, when the
mediator is the tensor particle. In this figure $\beta$ is defined as
$\beta=\lambda_2^\prime/\lambda_2$. Similar to spin--0 unparticle mediator
case, we set $\lambda_2=1$ in numerical calculations. It follows from this
figure that, when the coupling constants of two gluon system with perpendicular
and parallel spin orientations are equal, the branching ratio of the
spin--perpendicular configuration exceeds the spin--parallel configuration
of the two--gluon system up to $d_{\cal U}=1.15$.
  
Note that all above--presented results are obtained at $\Lambda_{\cal U} =
1~TeV$. In this connection the question, how restrictions on $d_{\cal U}$
depend on the cut--off parameter $\Lambda_{\cal U}$, should be considered.
In other words, at which parametric region of $d_{\cal U}$ and
$\Lambda_{\cal U}$ the branching ratio is larger than $10^{-9}$. In order to
answer this question, we present in Figs. (3) and (4) the parametric plot of the
branching ratio with respect to $d_{\cal U}$ and $\Lambda_{\cal U}$ which
gives ${\cal B}=10^{-9}$, for the $t \rar c g g$ decay, at fixed the values of
$C_S=C_P=10^{-1},10^{-2}$ and $C=10^{-2}$, in the presence of the
scalar operator. The region on the right side of each curve should be
excluded, since ${\cal B} <  10^{-9}$ in this domain. 
We observe that stringent constraints due to $d_{\cal U}$
and $\Lambda_{\cal U}$ are obtained for the $C_P=C_S=10^{-2}$ case.

Figs. (5)--(7) depict the the same analysis for the tensor
operator. It follows from these figures that the branching ratio is
reachable to be investigated up to $\Lambda_{\cal U}=10~TeV$
and up to $d_{\cal U}=1.5$.

In conclusion, we analyze the rare $t \rar c g g$ decay, that can exist 
at tree level in unparticle physics. Note
that these decays can take place only at loop level in the SM. For this
reason the branching ratio of these decays in unparticle physics can exceed
the ones predicted by the SM. The experimental measurement of the branching
ratios larger than $10^{-9}$ can give valuable information about the
existence of the new physics beyond the SM, in particular, about the
unparticle physics.

\newpage

\section*{Figure captions}
{\bf Fig. 1} The dependence of the branching ratio of the $t \rar cgg$ decay
on $d_{\cal U}$, at the values $C_P=C_S=10^{-2}$ of the t--c unparticle 
coupling constants, at $C=10^{-2}$ of the cut--off parameter, and at 
$\Lambda_{\cal U} = 1~TeV$, when the scalar unparticle is the mediator. \\ \\
{\bf Fig. 2}  The same as in Fig. (1), but when the tensor unparticle is the
mediator. \\ \\
{\bf Fig. 3} The parametric plot of the dependence of $\Lambda_{\cal U}$ on
the scaling parameter $d_{\cal U}$ at $C=10^{-2}$ and $C_P=C_S=10^{-1}$, when
branching ratio for the $t \rar c g g$ decay ${\cal B}(t \rar c g g)=1.2
\times 10^{-9}$, and when the scalar unparticle is the mediator. \\ \\
{\bf Fig. 4} The same as in Fig. (3), but at $C_P=C_S=10^{-2}$. \\ \\
{\bf Fig. 5} The same as in Fig. (3), but at $C=10^{-1}$, in the presence of
tensor unparticles. \\ \\
{\bf Fig. 6} The same as in Fig. (5), but at $C=10^{-2}$. \\ \\
{\bf Fig. 7} The same as in Fig. (5), but at $C=10^{-3}$. 

\newpage

\begin{figure}  
\vskip 4.0 cm   
    \includegraphics{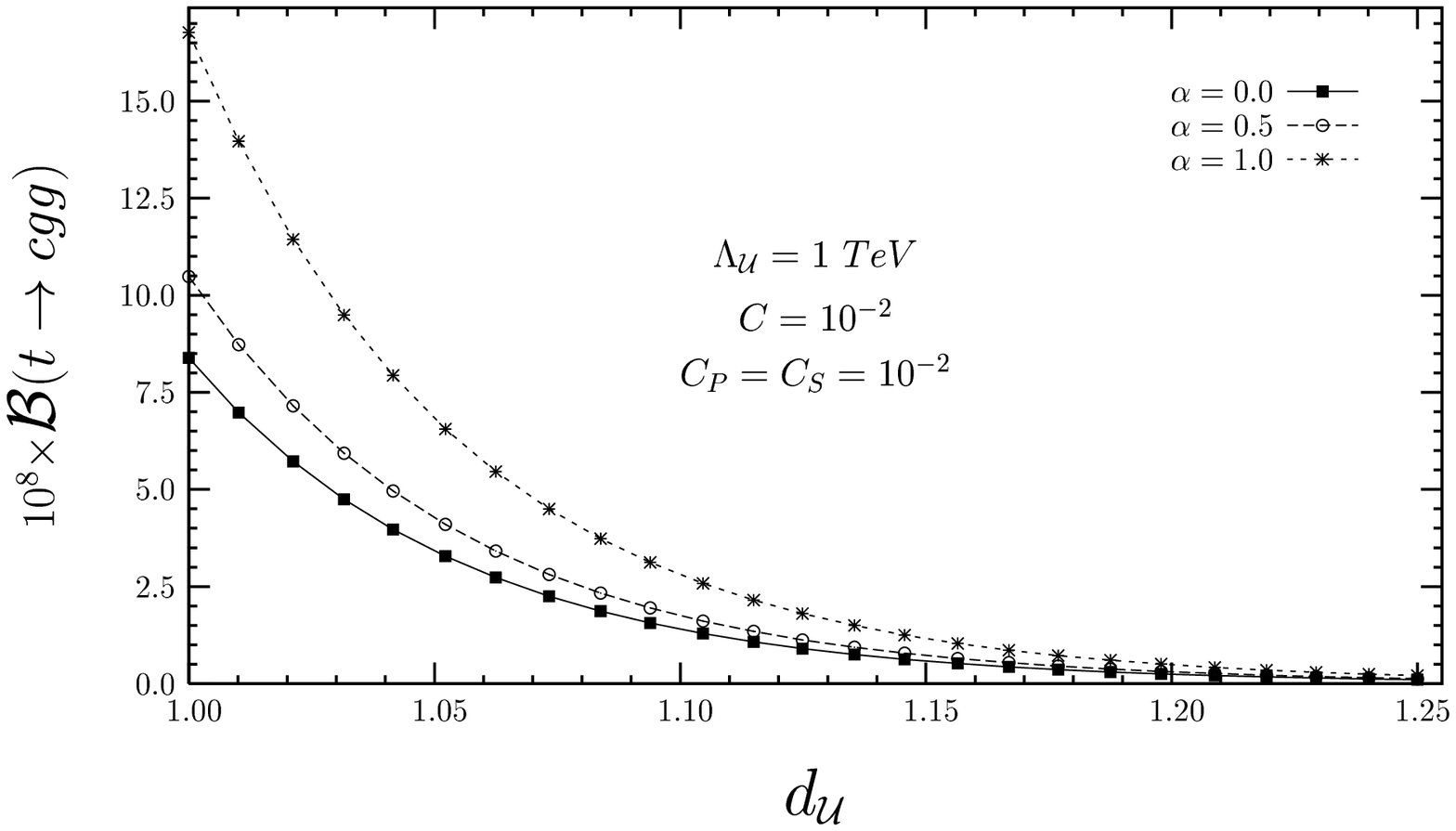}
\vskip 6.3 cm
\caption{}
\end{figure}

\begin{figure}  
\vskip 4.0 cm
    \includegraphics{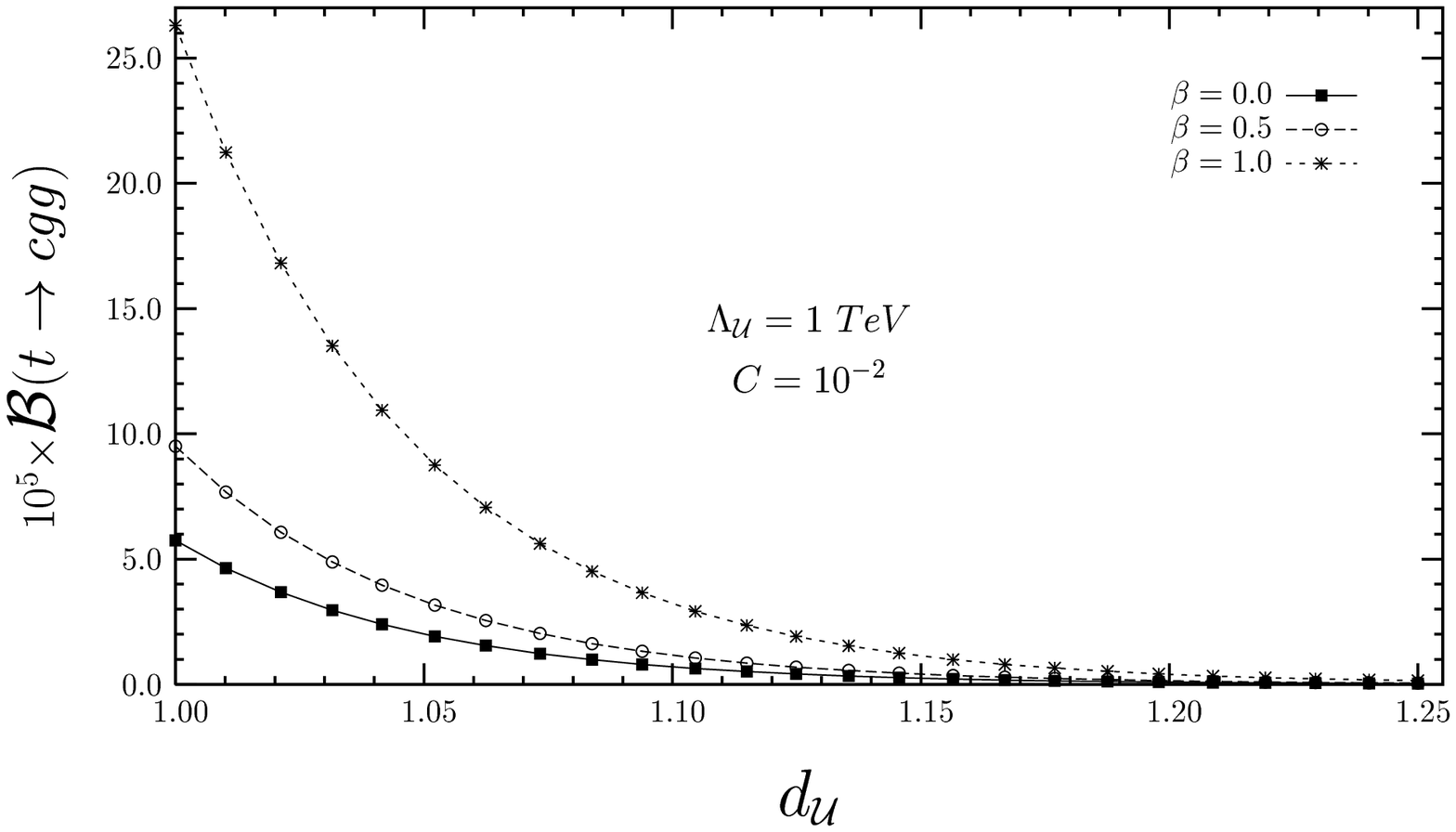}
\vskip 6.3 cm
\caption{}
\end{figure}

\begin{figure}
\vskip 4.0 cm
    \includegraphics{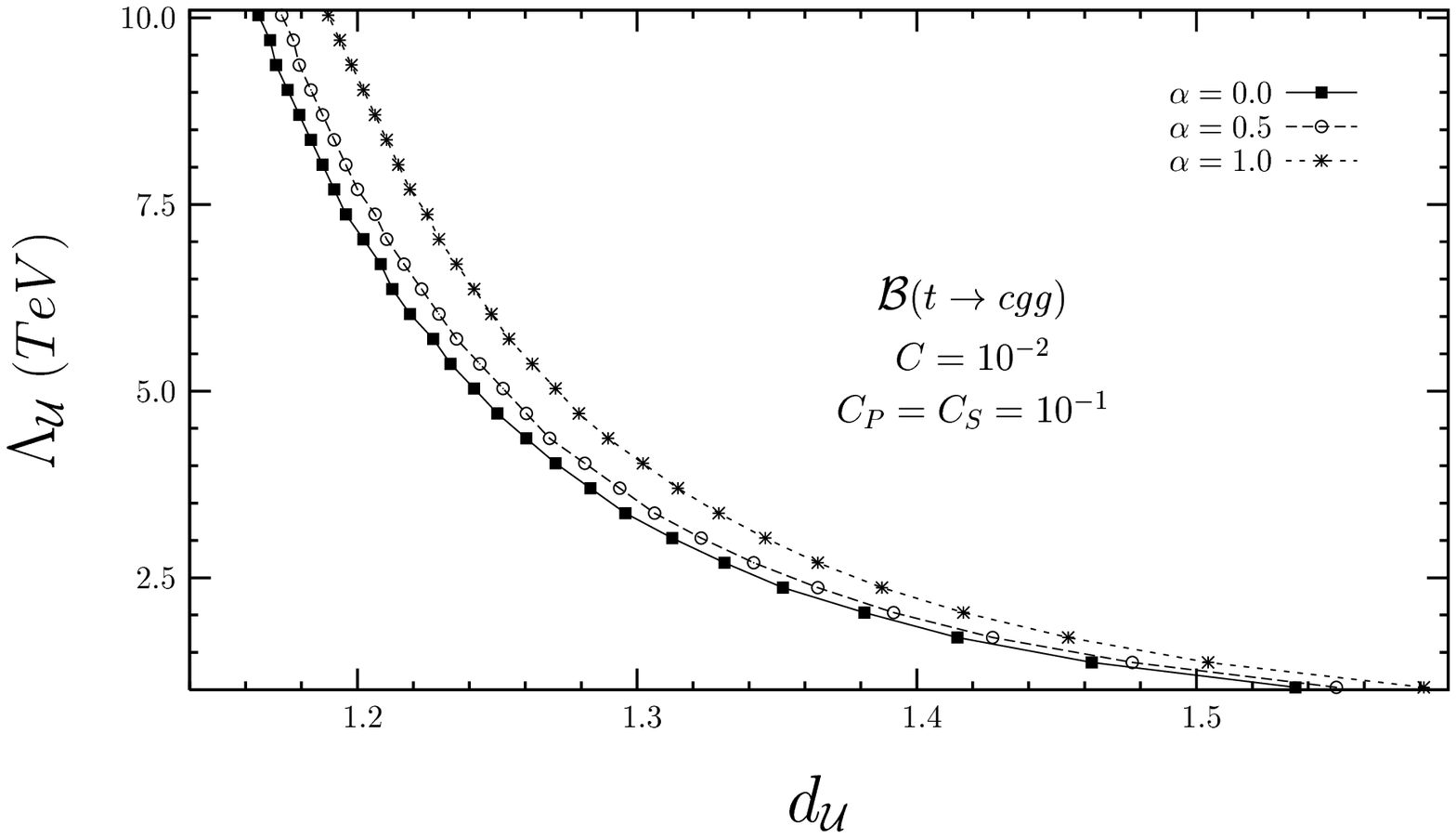}
\vskip 6.3 cm
\caption{}      
\end{figure}

\begin{figure}
\vskip 4.0 cm
    \includegraphics{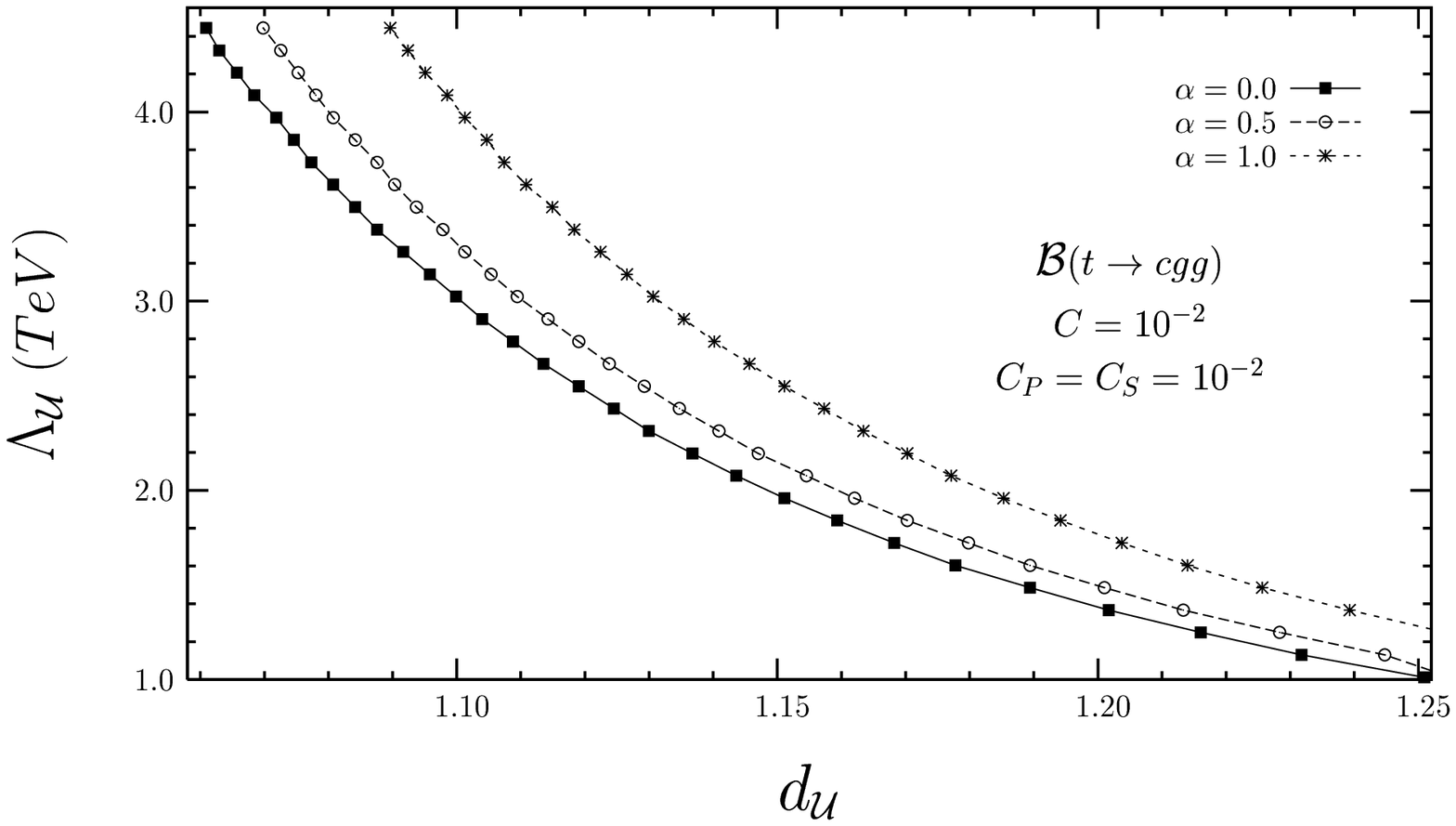}
\vskip 6.3 cm
\caption{}      
\end{figure}

\begin{figure}
\vskip 4.0 cm
    \includegraphics{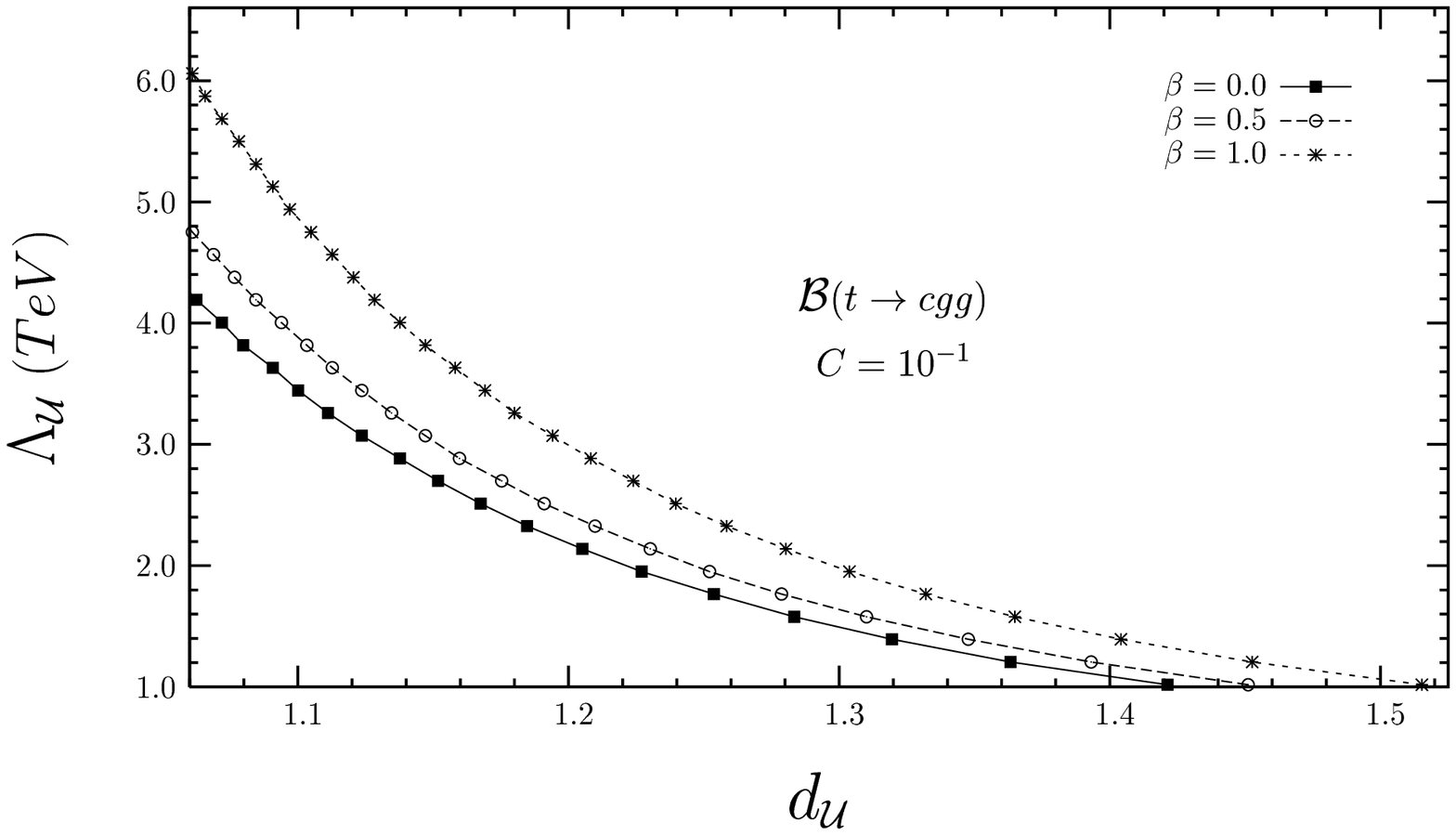}
\vskip 6.3 cm
\caption{}      
\end{figure}

\begin{figure}
\vskip 4.0 cm
    \includegraphics{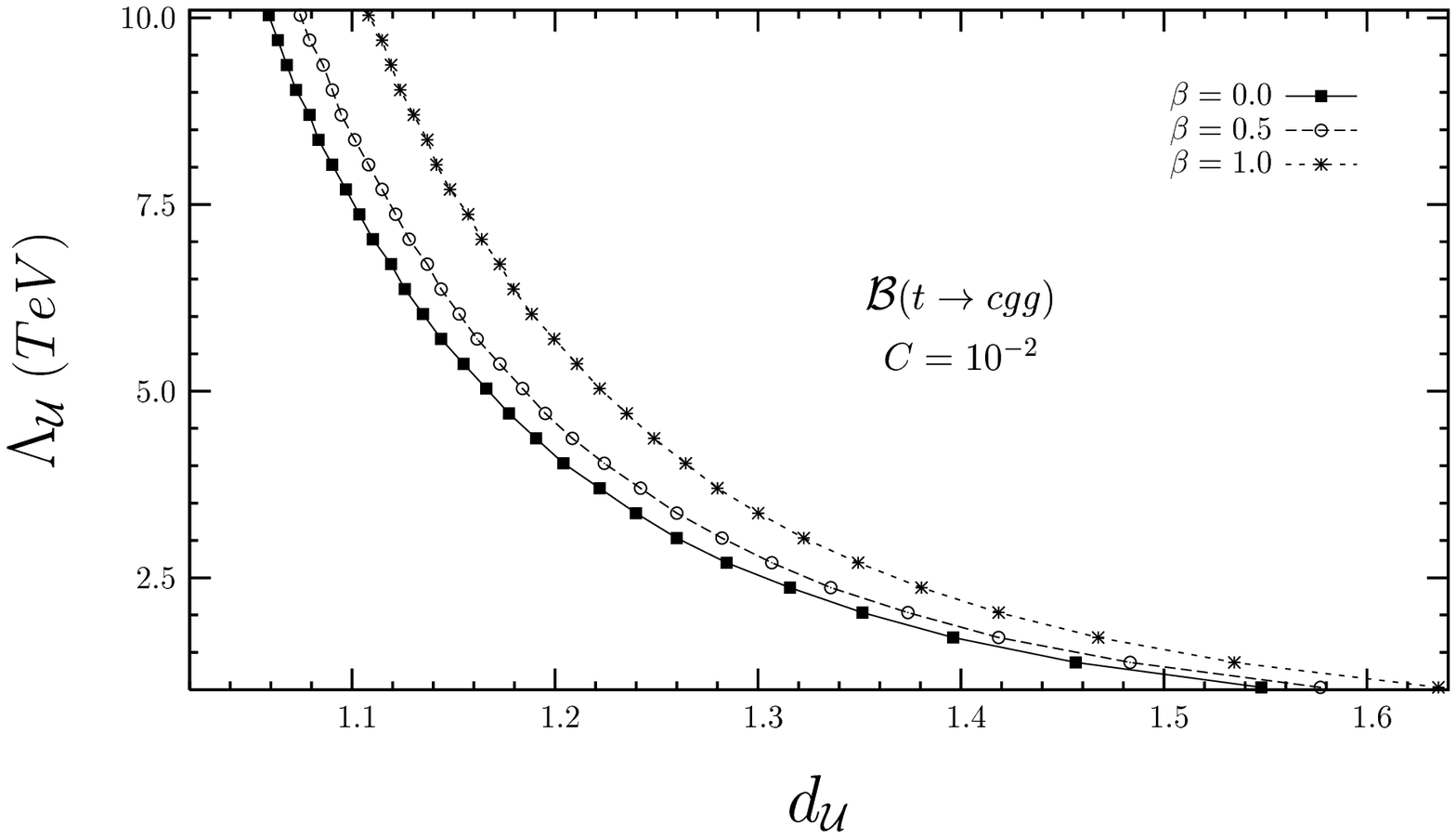}
\vskip 6.3 cm
\caption{}      
\end{figure}

\begin{figure}  
\vskip 4.0 cm   
    \includegraphics{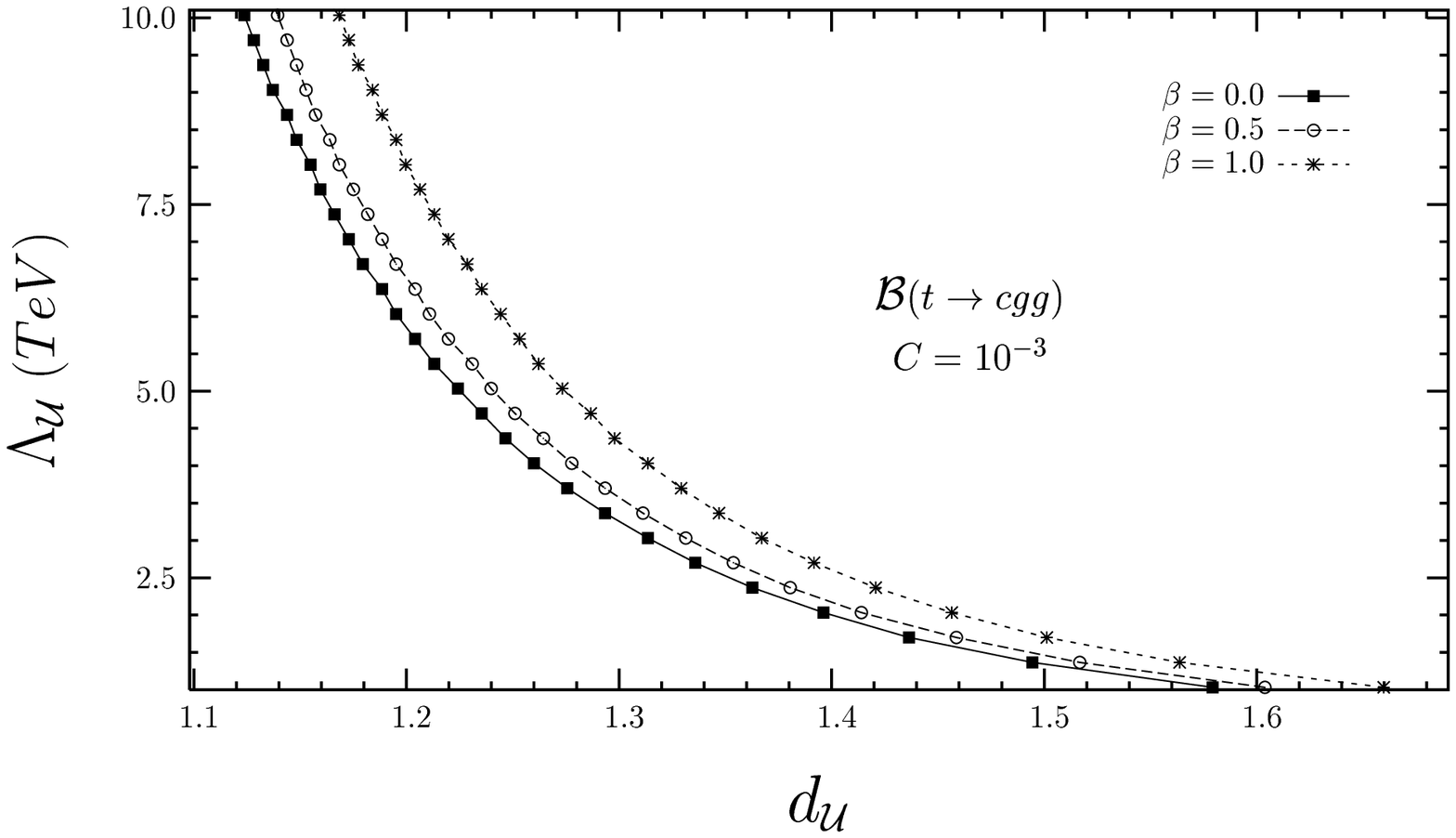}
\vskip 6.3 cm
\caption{}
\end{figure}


\newpage

\begin{thebibliography}{99}

\bibitem{R9101}  
 D. Brown,
  in "Proc. of Lepton--Photon 2007 Symposium".

\bibitem{R9102}
 M. Nakao,
  in "Proc. of Lepton--Photon 2007 Symposium".

\bibitem{R9103}
 J. Witch,
  (BELLE Collaboration),
  arXiv:  0711.0271 [hep--ex]. 

\bibitem{R9104}
 E. Barberio {\it et al.},
  arXiv:  0704.3575 [hep--ex].

\bibitem{R9105}
 G. Eliam, J. L. Hewett and A. Soni,
  Phys. Rev. D {\bf 44}, 1473 (1991);
 {\it Erratum}, D {\bf 59}, 039901 (1999) 

\bibitem{R9106} 
 B. Mele, S. Petrarca and A. Sodda,
  Phys. Lett. B {\bf 435}, 401 (1998).

\bibitem{R9107}
 J. Carvalho, N. Castro, A. Onofre and F. Velosco,
  (ATLAS Collaboration), Atlas Internal Note,
  ATL--PHYS--PUB--2005--009.

\bibitem{R9108}
 M. Cobal, 
  AIP Conf. Proc. {\bf 753}, 234 (2005).

\bibitem{R9109}
 M. Frank and I. Turan,
  Phys. Rev. D {\bf 72}, 035008 (2005).

\bibitem{R9110}
 E. Jenkins,
  Phys. Rev. D {\bf 56}, 458 (1997).

\bibitem{R9111}
 G. Altarelli, L. Conti and V. Lubicz,
  Phys. Lett. B {\bf 502}, 125 (2001).

\bibitem{R9112}
 S. Bar--Shalom, G. Eliam, M. Frank and I. Turan,
  Phys. Rev. D {\bf 72}, 055018 (2005).

\bibitem{R9113}
 S. Bar--Shalom, G. Eliam, A. Soni and J. Wudka,
  Phys. Rev. D {\bf 57}, 2957 (1998).     

\bibitem{R9114}
 H. Georgi,
  Phys. Rev. Lett. {\bf 98}, 221601 (2007);
 H. Georgi,
  Phys. Lett. B {\bf 650}, 275 (2007).

\bibitem{R9115}
 Y. Liao,
  Phys. Rev.  D {\bf 76}, 056006 (2007);
 M. Luo and G. Zhu,
  arXiv:  0704.3532 [hep-ph];
 I. Sahin and B. Sahin,
  arXiv:  0711.1665 [hep-ph];
 O. Cakir and K. O. Ozansoy,
  arXiv:  0710.5773 [hep-ph];
 E. O. Iltan,
  arXiv:  0710.2677 [hep-ph].

\bibitem{R9116}
 K. Cheung, W. Y. Keung and T. C. Yuan,
  Phys. Rev. Lett.  {\bf 99}, 051803 (2007);
 K. Cheung, W. Y. Keung and T. C. Yuan,
  Phys. Rev. D {\bf 76}, 055003 (2007).

\bibitem{R9117}
  S. L. Chen, X. G. He and H. C. Tsai,
  JHEP {\bf 0711}, 010 (2007);
  M. Luo, W. Wu and G. Zhu,
  Phys. Lett. B {\bf 659}, 359 (2008);
  N. Greiner,
  Phys.  Lett. B {\bf 653}, 75 (2007);
  P. Mathews and V. Ravindran,
  Phys. Lett. B {\bf 657}, 198 (2007);
  M. C. Kumar, P. Mathews, V. Ravindran and A. Tripathi,
  Phys. Rev. D {\bf 77}, 055013 (2008);
  K. Cheung, C. S. Li and T. C. Yuan,
  Phys.  Lett. B {\bf 662}, 438 (2008);
  P. J. Fox, A. Rajaraman and Y. Shirman,
  Phys. Rev. D {\bf 76}, 075004 (2007);
  A. T. Alan and N. K. Pak,
  arXiv: 0708.3802 [hep-ph];
  A. T. Alan, N. K. Pak and A. Senol,
  arXiv: 0710.4239 [hep-ph];
  O. Cakir and K. O. Ozansoy,
  arXiv: 0710.5773 [hep-ph];
  I. Sahin,
  arXiv: 0802.2818 [hep-ph];
 B. Sahin,
  arXiv: 0802.1937 [hep-ph];
 H. F. Li, H. l. Li, Z. G. Si and Z. J. Yang,
  arXiv: 0802.0236 [hep-ph];
 V. Barger, Y. Gao, W. Y. Keung, D. Marfatia and V. N. Senoguz,
  Phys.  Lett. B {\bf 661}, 276 (2008);
  C. F. Chang, K. Cheung and T. C. Yuan,
  arXiv: 0801.2843 [hep-ph];
 K. Cheung, T. W. Kephart, W. Y. Keung and T. C. Yuan,
  Phys.  Lett. B {\bf 662}, 436 (2008);
  O. Cakir and K. O. Ozansoy,
  arXiv: 0712.3814 [hep-ph];
 T. Kikuchi, N. Okada and M. Takeuchi,
  arXiv:  0801.0018 [hep-ph];
 T. Kikuchi and N. Okada,
  Phys.  Lett. B {\bf 661}, 360 (2008);
 J. R. Mureika,
  Phys. Lett. B {\bf 660}, 561 (2008);
  K. Huitu and S. K. Rai,
  Phys. Rev. D {\bf 77}, 035015 (2008);
 A. T. Alan, N. K. Pak and A. Senol,
  arXiv: 0710.4239 [hep-ph];
  A. T. Alan,
  arXiv: 0711.3272 [hep-ph].

\bibitem{R9118}
 C. H. Chen and C. Q. Geng,
  Phys. Rev. D {\bf 76}, 115003 (2007);
 C. H. Chen and C. Q. Geng,
  Phys. Rev. D {\bf 76}, 036007 (2007);
 R. Mohanta and A. K. Giri,
  arXiv: 0711.3516 [hep-ph];
 V. Bashiry,
  arXiv: 0801.1490 [hep-ph].

\bibitem{R9119} 
 G. J. Ding and M. L. Yan,
  Phys. Rev. D {\bf 76}, 075005 (2007).

\bibitem{R9120}
 T. M. Aliev, A. S. Cornell and N. Gaur,
  Phys. Lett. B {\bf 657}, 77 (2007);
 C. D. Lu, W. Wang and Y. M. Wang,
  Phys. Rev. D {\bf 76}, 077701 (2007);
 A. Hektor, Y. Kajiyama and K. Kannike,
  arXiv: 0802.4015 [hep-ph];
 E. O. Iltan,
  arXiv: 0802.1277 [hep-ph;] and  arXiv: 0801.0301 [hep-ph]; 
 G. J. Ding and M. L. Yan,
  Phys. Rev. D {\bf 77}, 014005 (2008).

\bibitem{R9121}
 X. Q. Li and Z. T. Wei,
  Phys. Lett. B {\bf 651}, 380 (2007);
 T. M. Aliev, A. S. Cornell and N. Gaur,
  JHEP {\bf 0707}, 072 (2007);
 R. Mohanta and A. K. Giri,
  Phys. Rev. D {\bf 76}, 075015 (2007); 
  Phys. Lett. B {\bf 660}, 376 (2008);
 A. Lenz,
  Phys. Rev. D {\bf 76} (2007) 065006;
 M. J. Aslam and C. D. Lu,
  arXiv: 0802.0739 [hep-ph];
 C. H. Chen, C. S. Kim and Y. W. Yoon,
  arXiv: 0801.0895 [hep-ph];
 Y. f. Wu and D. X. Zhang,
  arXiv: 0712.3923 [hep-ph];
 S. L. Chen, X. G. He, X. Q. Li, H. C. Tsai and Z. T. Wei,
  arXiv: 0710.3663 [hep-ph];
 T. M. Aliev and M. Savci,
  Phys. Lett. B {\bf 662}, 165 (2008).

\bibitem{R9122}
 X. Q. Li, Y. Liu and Z. T. Wei,
  arXiv: 0707.2285 [hep-ph];
 S. Zhou,
  arXiv: 0706.0302 [hep-ph];
 D. Montanino, M. Picariello and J. Pulido,
  arXiv: 0801.2643 [hep-ph];
 S. Dutta and A. Goyal,
  arXiv: 0801.2143 [hep-ph];
 A. B. Balantekin and K. O. Ozansoy,
  Phys. Rev. D {\bf 76}, 095014 (2007).

\bibitem{R9123}
 L. Anchordoqui and H. Goldberg,
  arXiv: 0709.0678 [hep-ph].

\bibitem{R9124}
 X. G. He and S. Pakvasa,
  arXiv: 0801.0189 [hep-ph].

\bibitem{R9125}
 H. Davoudiasl,
  Phys. Rev. Lett. {\bf 99}, 141301 (2007);
 J. McDonald,
  arXiv: 0709.2350 [hep-ph].

\bibitem{R9126}
 S. Hannestad, G. Raffelt and Y. Y. Y. Wong,
  arXiv: 0708.1404 [hep-ph];
 P. K. Das,
  arXiv: 0708.2812 [hep-ph];
 I. Lewis,
  arXiv: 0710.4147 [hep-ph];
 H. Collins and R. Holman,
  arXiv: 0802.4416 [hep-ph];
 T. Kikuchi and N. Okada,
  arXiv: 0711.1506 [hep-ph];
 S. L. Chen, X. G. He, X. P. Hu and Y. Liao,
  arXiv: 0710.5129 [hep-ph];
 G. L. Alberghi, A. Y. Kamenshchik, A. Tronconi, G. P. Vacca and G. Venturi,
  arXiv: 0710.4275 [hep-th].

\bibitem{R9127}
 A. Freitas and D. Wyler,
  arXiv: 0708.4339 [hep-ph].

\bibitem{R9128}
 K. Cheung, W. Y. Keung and T. C. Yuan,
  arXiv: 0706.3155 [hep--ph].

\bibitem{R9129}
 B. Grinstein, K. Intriligator, I. Z. Rothstein
  arXiv:0801.1140 [hep-ph].

\bibitem{R9130}
  H. Simma and D. Wyler,
  Nucl. Phys. B {\bf 344}, 283 (1990).

\bibitem{R9131}
  T. D. Lee and M. Nauenberg,
  Phys. Rev. {\bf 133}, B1549 (1964). 

\bibitem{R9132}
  G. Sterman,
  Phys. Rev. D {\bf 14}, 2123 (1976).


\end{thebibliography}
\end{document}